\documentclass[conference]{IEEEtran}
\IEEEoverridecommandlockouts
\usepackage{cite}
\usepackage{amsmath,amssymb,amsfonts}
\usepackage{algorithmic}
\usepackage{graphicx}
\usepackage{textcomp}
\usepackage{xcolor}
\usepackage{pifont}
\usepackage{multirow}
\usepackage{multicol}
\usepackage{subfig}
\usepackage{booktabs}
\usepackage{cellspace}

\def\BibTeX{{\rm B\kern-.05em{\sc i\kern-.025em b}\kern-.08em
    T\kern-.1667em\lower.7ex\hbox{E}\kern-.125emX}}
\begin{document}

\title{Blockchain-based Federated Learning for Decentralized Energy Management Systems\\

}

\author{\IEEEauthorblockN{Abdulrezzak Zekiye, Öznur Özkasap}
\IEEEauthorblockA{{Department of Computer Engineering}, {Koç University}\\
Istanbul, Türkiye \\
\{azakieh22, oozkasap\}@ku.edu.tr}
}

\maketitle

\begin{abstract}
The Internet of Energy (IoE) is a distributed paradigm that leverages smart networks and distributed system technologies to enable decentralized energy systems. In contrast to the traditional centralized energy systems, distributed Energy Internet systems comprise multiple components and communication requirements that demand innovative technologies for decentralization, reliability, efficiency, and security. Recent advances in blockchain architectures, smart contracts, and distributed federated learning technologies have opened up new opportunities for realizing decentralized Energy Internet services.
In this paper, we present a comprehensive analysis and classification of state-of-the-art solutions that employ blockchain, smart contracts, and federated learning for the IoE domains. Specifically, we identify four representative system models and discuss their key aspects. These models demonstrate the diverse ways in which blockchain, smart contracts, and federated learning can be integrated to support the main domains of IoE, namely distributed energy trading and sharing, smart microgrid energy networks, and electric and connected vehicle management.
Furthermore, we provide a detailed comparison of the different levels of decentralization, the advantages of federated learning, and the benefits of using blockchain for the IoE systems. Additionally, we identify open issues and areas for future research for integrating federated learning and blockchain in the Internet of Energy domains.
\end{abstract}


\section{Introduction}
The Internet of Energy, also known as the Energy Internet, is a cutting-edge distributed approach that merges smart networks and internet technology \cite{wang2018ieee, hussain2019emerging}. Unlike conventional centralized energy systems, Energy Internet systems are composed of multiple components with varying communication requirements, necessitating innovative technologies to guarantee dependability, efficiency, and security \cite{wang2017survey}. Decentralized technologies such as blockchain, smart contracts, and distributed federated learning present promising prospects for enhancing Energy Internet systems.

In this study, we identify and examine three primary domains of the Internet of Energy systems, namely: (i) Distributed energy trading and sharing, (ii) Smart microgrid energy networks, and (iii) Electric and connected vehicle management.

\textbf{Distributed Energy Trading and Sharing (DETS):} refers to the exchange of energy between individuals or organizations using distributed energy resources, such as solar panels, battery storage systems, and wind turbines. This type of trading and sharing allows for the decentralized generation and distribution of energy, enabling participants to buy and sell excess energy generated from their own renewable energy systems \cite{soto2021peer}. Trading typically occurs in exchange for payment, while sharing is done in exchange for future benefits \cite{bouachir2022federatedgrids}. 

\textbf{Smart microgrid energy networks (SMEN):} A smart microgrid is a small-scale energy network that integrates renewable energy sources, such as solar and wind power, with traditional sources of electricity. It utilizes advanced control systems and technologies, such as smart meters and energy storage systems, to optimize energy generation, distribution, and consumption in a more efficient and sustainable way \cite{cecati2010overview}. We consider energy trading and sharing systems that involve inter-microgrid operations as falling under this domain. There exist studies on energy trading systems in microgrids, such as \cite{wang2017novel} where a novel system has been proposed to decentralize the electricity transaction mode of microgrids. The proposed system relies mainly on blockchain and continuous double action.

\textbf{Electric and connected vehicle management (ECVM):} This domain of IoE refers to the use of advanced technologies and techniques to optimize the operation and performance of electric vehicles (EVs) and connected vehicles (CVs) in the transportation sector. This domain can be divided into three main aspects: energy flow, data communication, and computation \cite{chen2020energy}.

Decentralized technologies such as novel blockchain architectures, smart contracts, and distributed federated learning offer opportunities for designing Energy Internet systems. In this study, one of the main contributions is analyzing and classifying the state-of-the-art solutions utilizing these technologies for the IoE domains. By means of the analysis, we identified the representative system models of the existing solutions. In the following paragraphs, we briefly describe the key terms related to the system models.

\textbf{Flat Blockchains:} Blockchain is a decentralized and distributed ledger technology that allows for the permanent storage of data. Its initial utilization was for the purpose of signing and timestamping documents\cite{haber1990time}. Subsequently, blockchain was proposed as the foundation of Bitcoin, a peer-to-peer cash system \cite{nakamoto2008bitcoin}. However, blockchain technology is not limited to financial applications, and researchers have explored its potential in diverse domains, such as health \cite{haleem2021blockchain} and supply chain management \cite{queiroz2019blockchain}. In \cite{zheng2018smart}, the researchers developed a smart grid power trading system that leveraged blockchain technology. Similarly, a framework for decentralized electric vehicle charging was proposed in \cite{su2018secure}, which utilized a permissioned blockchain as the underlying infrastructure. We refer to the traditional (non-hierarchical) blockchain architectures as \textit{flat blockchains}. 

\textbf{Hierarchical Blockchains:} Hierarchical blockchains have been introduced to solve the scalability problem in large-scale networks. In such blockchains, the nodes are organized
into levels. Each level can be a blockchain accessed by different nodes with different functionalities. For instance, in the system of \cite{sahoo2019hierarchical}, local transactions were stored in multiple local blockchains lying at a specific level. At the upper level, there is a separate blockchain responsible for storing partial views of the local blockchain at the lower level. There exists another structure of hierarchical blockchains in \cite{lu2020blockchain} that is called a hybrid blockchain. The structure had permissioned blockchain accessible by Road-Side Units, and local Directed Acyclic Graphs accessed by vehicles. We call such systems \textit{hierarchical blockchains} since the permissioned blockchain and local DAGs can be considered as levels in a hierarchical manner.

\textbf{Smart Contracts:} A smart contract is a type of computer code that runs over blockchain and can automatically enforce the terms of an agreement. It is designed to be self-executing, self-verifying, and secure against tampering \cite{mohanta2018overview}. This allows for the automation of digital contracts, which can help to streamline processes and reduce the need for intermediaries. \textcolor{black}{Usage of smart contracts varies widely. A well-known application of smart contracts is the creation of Non-Fungible Tokens (NFTs). Smart contracts have been also been used \cite{ali2020synergychain} to develop an adaptive model for prosumer grouping in peer-to-peer energy trading systems, where the grouping process is handled by a smart contract.}

\textbf{Federated Learning:} In federated learning \textcolor{black}{(FL)}, a central server initializes and distributes a machine-learning model to multiple devices, and they train the model on their own data. These devices send back their updated model parameters to the central server, which combines these updates and sends a new version of the model back to the devices. This process continues until the model achieves convergence, at which point the final version of the model is returned to the central server \cite{konevcny2016federated}. As stated in \cite{mammen2021federated}, there are different types of federated learning, including Horizontal FL, Vertical FL, Federated Transfer Learning, Cross-Silo FL, and Cross-Device FL. Federated learning has applications in various fields, such as healthcare, transportation, finance, and wireless communications\cite{mammen2021federated, zhang2021survey}.  In addition, researchers have also utilized federated learning in the Internet of Energy field \cite{bouachir2022federatedgrids}.

The contributions of this study are as follows:

\begin{itemize}

\item We conducted a thorough analysis of the current state-of-the-art solutions that integrate blockchain, smart contracts, and federated learning principles in the primary domains of Internet of Energy systems. Our study is the first, to the best of our knowledge, to perform such an analysis and classification in the Internet of Energy domains. Researchers in \cite{wu2018application, miglani2020blockchain} conducted reviews of the use of blockchain in the Internet of Energy field,  but did not consider the use of federated learning. In \cite{billah2022systematic}, both federated learning and blockchain were considered, but within the scope of the Internet of Vehicles and without providing representative system models. In \cite{ali2020cyberphysical},  a classification of blockchain-enabled energy trading systems into three models was proposed, but they focused solely on blockchain-enabled systems, particularly energy trading systems. 
In \cite{nguyen2021federated}, a general overview of the integration of blockchain with federated learning was presented, while also identifying some issues.
In the work \cite{banabilah2022federated}, researchers presented their efforts to integrate blockchain and federated learning, building upon previous works that incorporated both techniques. However, their primary focus was on federated learning and did not specifically address the use of these technologies in the context of the Internet of Energy. Additionally, they did not extract representative models. Table \ref{tab:previous} provides a comparison of our work with the related works. 

\item Based on the analysis and classification of the solutions and the consideration of the blockchain architecture and aggregation approach, we have identified four representative system models that demonstrate the potential for integrating blockchain and federated learning technologies in the Internet of Energy. These system models can provide a useful guideline for future research in this domain.

\item We have defined four levels of decentralization and have subsequently mapped them to the extracted system models.

\item One of the system models, Federated Learning Hierarchical Blockchain-based Systems with
Decentralized Aggregation, has the potential to offer improved scalability and reduced latency compared to the other system models. This is mainly due to its hierarchical structure and lack of a single point of failure.

\item We have discussed several open issues and potential areas for future research and development in the use of federated learning and blockchain-based systems in the Internet of Energy domain.

\end{itemize}

The paper is organized as follows. Section II offers an overview of the representative systems models of the state-of-the-art solutions. Sections III and IV provide a detailed description of the four system models and a comprehensive discussion of the solutions mapped to each system model. In Section V, a comparison of the system models and key findings are presented. Finally, Section VI identifies open issues and future research directions.

\begin{table}[htbp]
\caption{Comparison with related works}
\begin{center}
\begin{tabular}{|c|c|c|c|c|}
\hline
\textbf{Ref} & \parbox[c]{1.5cm}{\textbf{Blockchain}} & \parbox[c]{1.5cm}{\textbf{Federated Learning}} & \parbox[c]{1.5cm}{\textbf{Internet of Energy}} & \parbox[c]{1.5cm}{\textbf{Extracted System Models}}  \\
\hline
\cite{wu2018application} & \ding{51} & X & Partially & X   \\
\hline
 \cite{miglani2020blockchain} & \ding{51} & X & \ding{51} & X   \\
 \hline
  \cite{billah2022systematic} & \ding{51} & \ding{51} & Partially & X  \\
 \hline
\cite{ali2020cyberphysical} & \ding{51} & X & Partially & \ding{51}  \\
 \hline
\cite{nguyen2021federated} & \ding{51} & \ding{51} & X & X  \\
\hline

\cite{banabilah2022federated} & \ding{51} & \ding{51} & Partially & X  \\
\hline
 Ours & \ding{51} & \ding{51} & \ding{51} & \ding{51}  \\
\hline

\end{tabular}
\label{tab:previous}
\end{center}
\end{table}

\section{CLASSIFICATION AND OVERVIEW OF THE SYSTEM MODELS}

\begin{figure}[!ht]
	\centering
	\subfloat[]{\label{subfig:system1} \includegraphics[width=0.43\textwidth]{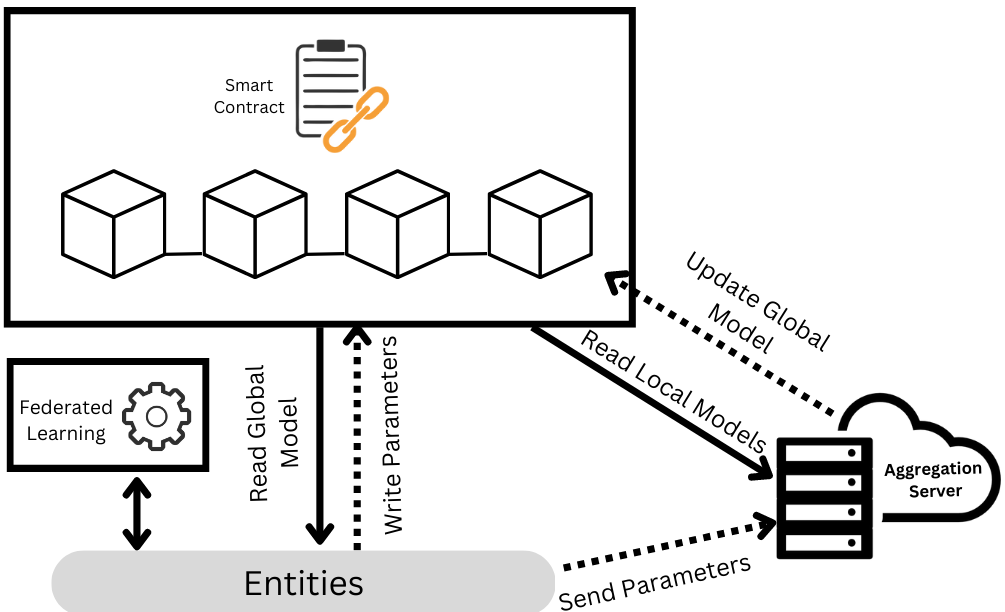}}  \vspace{5pt}
	\subfloat[]{\label{subfig:system2} \includegraphics[width=0.31\textwidth]{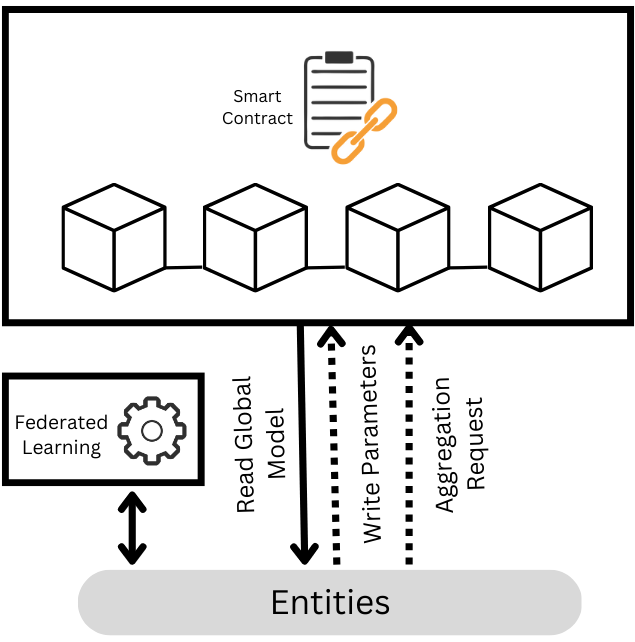}} \\
    \caption{Flat Blockchain and Federated Learning-Based Models in the Internet of
Energy: (a) Federated Learning and Flat Blockchain-based Systems with
Centralized Aggregation, (b) Federated Learning and Flat Blockchain-based Systems with Decentralized Aggregation. In (a), the entities either send local models to the server for aggregation or they write them to the blockchain where the server can access them. In (b), entities might request for aggregating the model, or the aggregation will be done automatically by the smart contract.}
    \label{fig:sys1and2}
\end{figure}

\begin{figure}[!ht]
	\centering
	\subfloat[]{\label{subfig:system3} \includegraphics[width=0.43\textwidth]{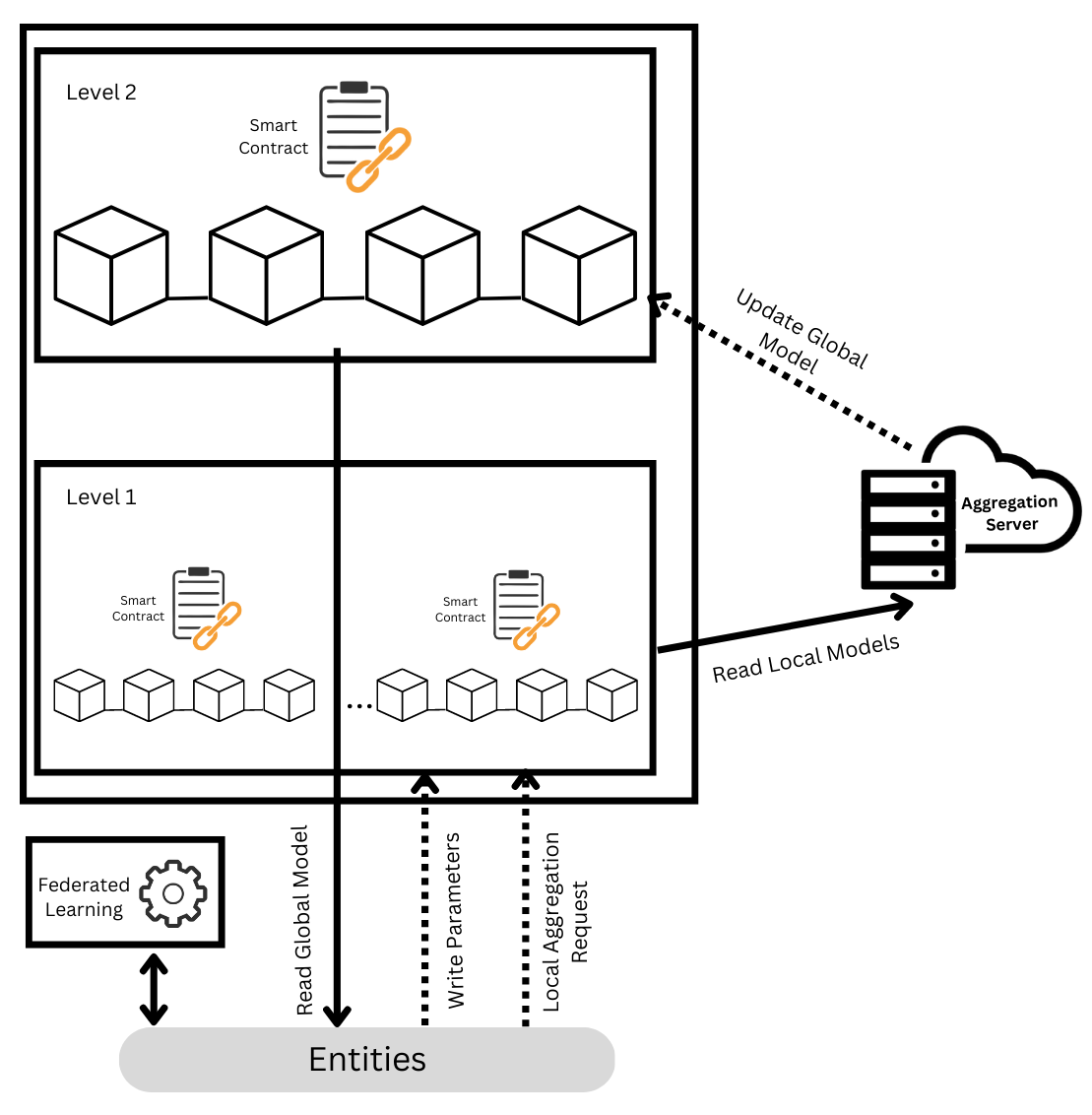}}  \vspace{10pt}
	\subfloat[]{\label{subfig:system4} \includegraphics[width=0.33\textwidth]{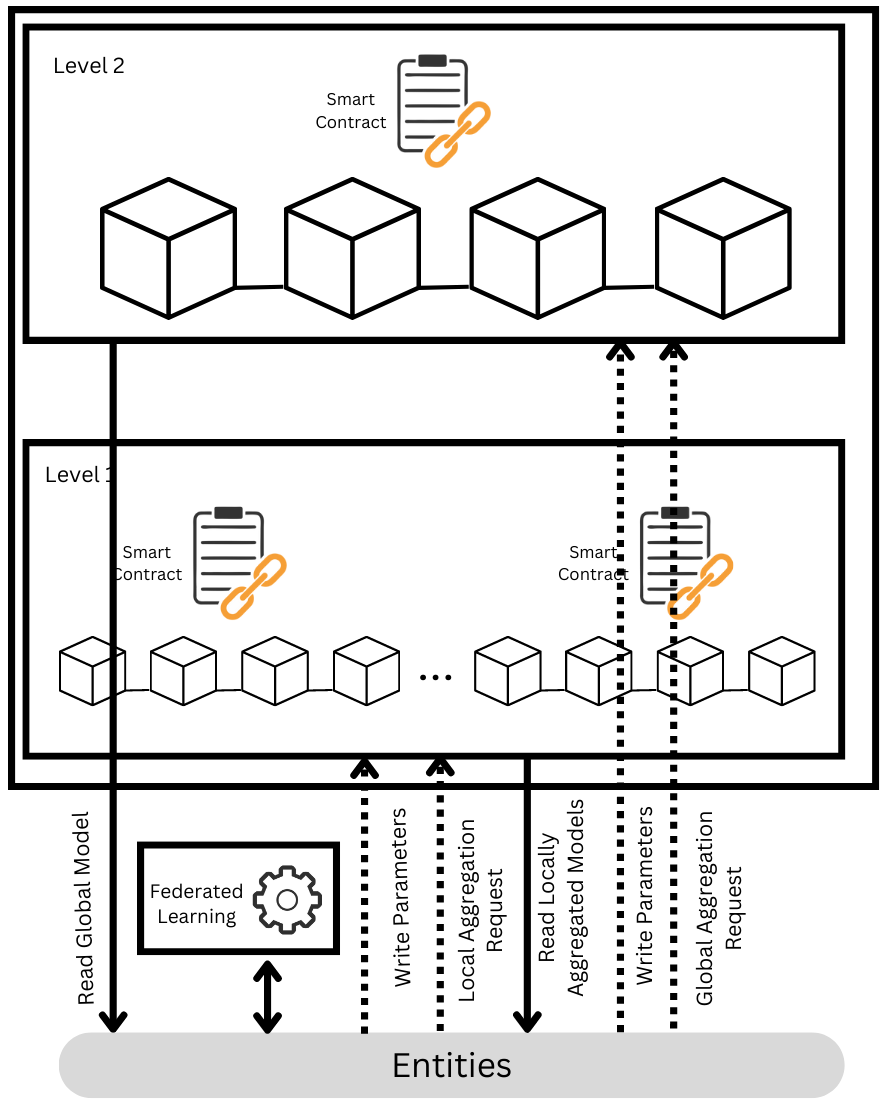}} \\
    \caption{Hierarchical Blockchain and Federated Learning-Based Models in
Internet of Energy (a) Federated Learning and Hierarchical Blockchain-based Systems with Centralized Aggregation, (b) Federated Learning and
Hierarchical Blockchain-based System with Decentralized Aggregation. Local aggregation requests might be done by entities in (a) and (b). In (b), global aggregation requests might exist.}
    \label{fig:sys3and4}
\end{figure}

With a focus on the utilization of both blockchain and federated learning in the Internet of Energy field, we identify four main system models by considering the type of blockchain and the aggregation model. Figures \ref{fig:sys1and2} and \ref{fig:sys3and4} show an illustration of the flat (non-hierarchical) blockchain and FL system models, and the hierarchical blockchain and FL system models respectively.

Entities in these systems depend on the domain in which the system model was found. Distributed energy trading and sharing domain has consumers, prosumers, utility grids,
stations, and servers as entities. While consumers use energy, prosumers both produce and consume energy. In the smart microgrid energy networks domain, smart grids, in addition to the entities in Distributed energy trading and sharing domain, are part of the system. Finally, Unmanned Aerial Vehicles (UAVs), Unmanned Ground Vehicles (UGVs), along with standard vehicles, Road Side Units (RSUs), Charging Stations, Fogs, and Servers are the key entities in the Electric and connected vehicle management domain.

It is important to state that in most cases, not all entities can do all the available operations. In addition, only the smart contracts in a blockchain can be used; while in some works, the smart contract and the consensus protocol are utilized. In some models federated learning is not handled by the peer itself, but by another entity with computation power. For that reason, the symbol of federated learning (\parbox[c]{0.04\textwidth}{\centering \includegraphics[width=0.03\textwidth]{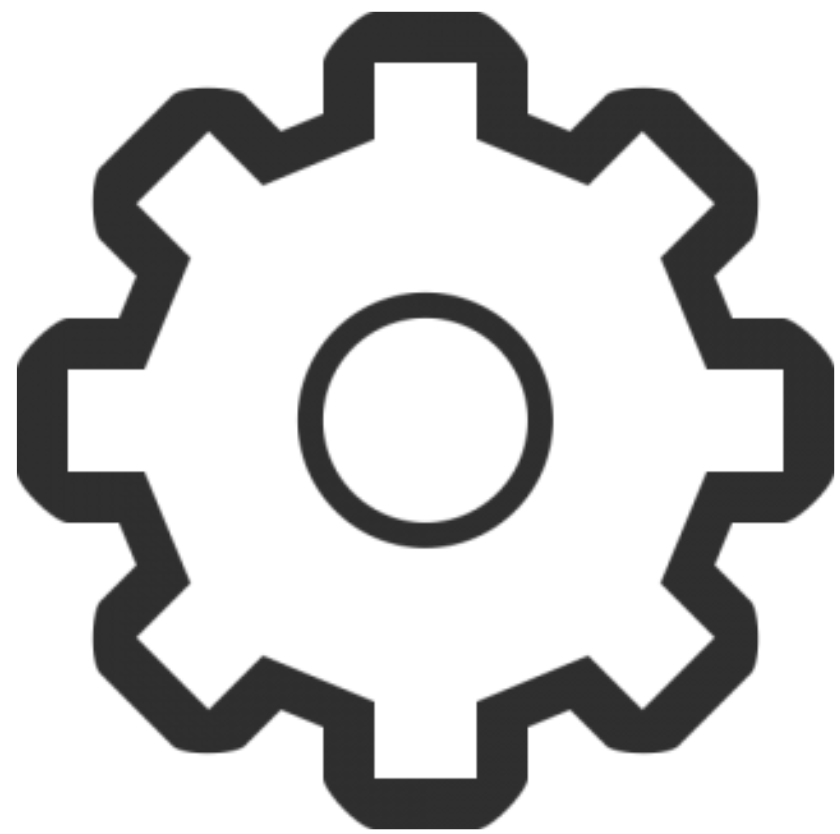}}) is separated from the entities.

Figure \ref{subfig:system2} shows the first system model, \textbf{Federated Learning and Flat Blockchain-based Systems with Centralized Aggregation (FLB-CA)}. In this model, entities can send the trained parameters to the aggregation server directly and then the aggregation server performs aggregation and stores the global model on the blockchain. Another possible way is that the entities upload local models to the blockchain, and then the server reads local models from the blockchain, aggregates, and writes the global model to the blockchain again. The initialization of the model is done by the server, as well as by writing it to the blockchain or broadcasting it to the related entities.

Figure \ref{subfig:system2} depicts the second system model, \textbf{Federated Learning and Flat Blockchain-based Systems with Decentralized Aggregation (FLB-DA)}. In this model, there
is no centralized server, thus the parameters are written to the blockchain by entities. Thanks to the smart contracts that run over the blockchain, some entities, or all of them in some cases, have the ability to perform the aggregation. The smart contract, upon being called by an entity, aggregates, and stores the global model in the blockchain. Initializing the model can be done by one of the entities or by an outside request.

The third system model, \textbf{Federated Learning and Hierarchical Blockchain-based Systems with Centralized Aggregation (FLHB-CA)}, is illustrated in Figure \ref{subfig:system3}. In this
system model, there is a centralized server that reads local models from the first (lower) level of the blockchain, does global aggregation, and updates the global model in the
second (upper) level of the blockchain. Entities read the global model from the upper level of the blockchain, do training using the local data, then upload the parameters to the first
level of the blockchain. Some entities can do local aggregations at the first level by calling a function from a smart contract. However, the final aggregation can be done by the server only as mentioned earlier.

\textbf{Federated Learning and Hierarchical Blockchain-based Systems with Decentralized Aggregation (FLHB-DA)} is the fourth extracted model as shown in Figure \ref{subfig:system4}. The difference from the third model (FLHB-CA) is that the centralized server does not exist here. Global aggregation is done on the second level of the blockchain thanks to smart contracts where some entities can make a request to do the global aggregation.

The first level of the hierarchical blockchain in Figures \ref{subfig:system3} and
\ref{subfig:system4} can be separate blockchains as in \cite{chai2020hierarchical}, local DAGs
\cite{lu2020blockchain}, or micro blocks \cite{ayaz2021blockchain} at specific entities, all of which are used for the same purpose, storing the local models or locally aggregated models in most of the works. The second level is a separate single blockchain, in the solutions of this category, which stores the aggregated global model. In the next two chapters, details of these system models, and the existing solutions that fall into each, are provided.

\section{FLAT BLOCKCHAIN WITH FL}

In the category of flat blockchain with federated learning, there exist two systems models, namely with centralized aggregation and with decentralized aggregation. Table \ref{tab:summary} shows the solutions \cite{bouachir2022federatedgrids}, \cite{otoum2022federated}, \cite{otoum2020blockchain}, \cite{he2021bift}, \cite{teimoori2022secure}, and \cite{aloqaily2021energy}, which those two models have been extracted from with the following information about each one of them: the aggregation type, level of decentralization, and whether the developed system operates only by using smart contracts or not. The definitions of decentralization levels are given in Section V.

\subsection{Federated Learning and Flat Blockchain-based Systems with Centralized Aggregation (FLB-CA)}

In such systems, a flat blockchain is used and a centralized server initializes the global model, asks for training it, and then does the aggregations to update the global model as illustrated in Fig. 1(a). The global model can be read by peers but cannot be updated by them. The updates are limited to the server. Since there is a single server to do the aggregation, it is a bottleneck in the system. If the server went down or got attacked, the whole system would malfunction. The solutions that are based on this system model (\cite{otoum2022federated}, \cite{otoum2020blockchain}, and \cite{aloqaily2021energy}) have the following characteristics.

In \cite{otoum2022federated}, the authors aimed to solve the insecurity and trustworthiness problems in the energy sector by using blockchain and federated learning for energy trading. The entities in their system were prosumers and consumers as peers, stations with computational power to run federated learning tasks, and a server to do the aggregation and initialize the model. Energy requests can be done by multiple nodes at the same time and each request is decomposed and a corresponding delivery plan is made by an AI model trained using federated learning. Federated learning was used to increase privacy by making peers not share raw data with the centralized server. In addition, the federated learning approach led to a higher delivery rate since the amount of data, the trained parameters, to be shared is less. The blockchain was used to save the transactions, and store peers’ devices’ capabilities. Moreover, the trained model is stored in the blockchain to be used later for speeding up the training process. They tested the model using 1000 nodes and three stations, then compared the model with non-federated learning models, namely vehicle-to-vehicle, vehicle-to-infrastructure, and vehicle-to-everything. The results show a higher delivery rate and minimal power consumption.

The study in \cite{otoum2020blockchain} used blockchain and federated learning to ensure network security and data privacy while training models in a vehicular network. A global model gets trained on the local data in the vehicles where the parameters get transferred to a server. The server passes the parameters trained on critical vehicles (e.g. ambulances, police cars, firefighting cars) through the blockchain to make sure that the critical vehicle is authentic. Parameters trained on the normal vehicles are passed by the server directly to fog. The fog does the aggregation using the directly passed parameters and the parameters that passed the blockchain and updates the global model. The usage of blockchain in this case was for increasing the privacy of critical vehicles. The idea of not passing all nodes through the blockchain reduced the latency. The system was tested on up to 100 vehicles with the MNIST dataset. The tools for simulation were MATLAB/Simulink, Contiki as an operating system, and Erlang nodes to represent the vehicles and the server. We can notice here that the training is handled by the peers themselves, while in \cite{otoum2022federated}, the training was done on stations.

In the work \cite{aloqaily2021energy}, the researchers aimed at providing continuous service to end devices using UAVs and UGVs. The structure they used is very similar to the one in \cite{otoum2020blockchain} where training is done on the peers, then instead of using a server to decide whether to pass the parameters through blockchain or not, a group leader is chosen. The group leader decides whether the data is sensitive and if so, passes it through the blockchain. If the data are not sensitive, it is then passed directly to the fog which performs the aggregation and updates the global model. The trained model aims to provide continuous service to end devices while considering the power availability of the UAVs.

\begin{table*}
\caption{{Summary of the state-of-the-art solutions mapped to the extracted system models}}

\centering
  \label{tab:summary}
  \begin{tabular}{|Sc|Sc|Sc|Sc|Sc|Sc|}
    \hline
    \parbox[c]{1cm}{\textbf{System \\ Model}}  & \textbf{Ref} &  \parbox[c]{2cm}{\textbf{Blockchain \\ Type}}  &  \parbox[c]{2cm}{\textbf{Aggregation \\ Type}} & \textbf{Domain} & \parbox[c]{3cm}{\textbf{Decentralization \\ Level}}    \\

    \hline
    \multirow{3}{*}{FLB-CA} & \cite{otoum2022federated} & \multirow{6}{*}{\parbox[c]{2cm}{Flat \\ Blockchain} } &
    \multirow{3}{*}{\parbox[c]{2cm}{Centralized \\ Aggregation}} & DETS & Medium  \\
    \cline{2-2}
    \cline{5-6}
      & \cite{otoum2020blockchain} & & & ECVM   & Medium  \\ 
    \cline{2-2}
    \cline{5-6}
    & \cite{aloqaily2021energy} & & & ECVM & Medium \\ 
    \cline{1-2}
    \cline{4-6}
    \multirow{3}{*}{FLB-DA} & \cite{bouachir2022federatedgrids} &   &
    \multirow{3}{*}{ \parbox[c]{2cm}{Decentralized \\ Aggregation}} & SMEN & Extreme  \\
    \cline{2-2}
    \cline{5-6}
      & \cite{he2021bift} & & & ECVM & Extreme \\ 
      \cline{2-2}
    \cline{5-6}
    & \cite{teimoori2022secure} & & & ECVM & High  \\ 
    \hline
    
    FLHB-CA & \cite{lu2020blockchain} & \multirow{3}{*}{\parbox[c]{2cm}{Hierarchical \\ Blockchain}} &
     \parbox[c]{2cm}{Centralized \\ Aggregation}   & ECVM & High \\
    \cline{1-2}
    \cline{4-6}
     \multirow{2}{*}{FLHB-DA} & \cite{chai2020hierarchical} &   &
     \multirow{2}{*}{\parbox[c]{2cm}{Decentralized \\ Aggregation}}  & ECVM & High \\
    \cline{2-2}
    \cline{5-6}
      & \cite{ayaz2021blockchain} & & & ECVM & High \\ 
  \bottomrule
\end{tabular}
\end{table*}

\subsection{Federated Learning and Flat Blockchain-based Systems with Decentralized Aggregation (FLB-DA)}

This model is utilized by the solutions in \cite{bouachir2022federatedgrids}, \cite{he2021bift}, and \cite{teimoori2022secure} where the flat blockchain is used. The difference is in the aggregation approach. While in the first model, a server was used for the aggregation, in this model the aggregation is done in a decentralized way thanks to the blockchain. Any peer or peers with specific privileges can do the aggregation by means of smart contracts, which aggregate the parameters and update the global model as well. This system model is represented in Fig. 1(b).

In \cite{bouachir2022federatedgrids}, an autonomous energy trading and sharing system inside and across microgrids, FederatedGrids, was proposed. In this system, energy sharing is trading but for future benefits instead of payments, and the proposed platform has consumers, prosumers, microgrids, and utility grids as entities. The platform uses flat blockchain together with federated learning. The blockchain supports payment, stores transactions, the global model, and information about entities, and hosts smart contracts. Smart contracts have many operations, and the most important ones are the prediction of energy production and demand in addition to the aggregation process. As a result, there is no server to do the aggregation and it can be done by any peer. For the evaluation, the Hourly Energy Consumption dataset was used and according to the obtained results, around 18\% improvement in cost for the energy consumers and 76\% improvement in load over utility grids were achieved.

The work in \cite{he2021bift} is similar to \cite{bouachir2022federatedgrids} considering the flat blockchain usage and handling the aggregations. The aim is to make secure and invulnerable traditional federated learning systems for the networks of connected and autonomous vehicles. Instead of sharing raw data, federated learning was used to share a trained model on the local data. The researchers coded a consortium (permissioned) blockchain and modified the consensus protocol to fit their needs. The blockchain plays a role in increasing security by being a consortium blockchain in addition to the usage of its consensus protocol. Data about the models are stored over the blockchain as well. Instead of storing the whole model on the blockchain, which might be huge, it was uploaded to an InterPlanetary File System (IPFS) and only the model hash was stored over the blockchain. The proposed system also supports sharing raw data when needed, like exact location, and it was done by uploading the data to IPFS and sharing the hash only over the blockchain.

In \cite{teimoori2022secure}, a charging station recommendation system was proposed. Federated learning was used to train the model without sharing raw data. Aggregation is handled by several cloudlets and those validator cloudlets are validated through a consortium blockchain. The blockchain also stores the trained parameters. Likewise, in this work, there is no single aggregator. However, the aggregation was not handled by the blockchain, i.e. smart contract, but it was done within the cloudlets.

\section{HIERARCHICAL BLOCKCHAIN WITH FL}

These models utilize hierarchical blockchain with federated learning. Two different aggregation methods, centralized and decentralized, are used. In Table \ref{tab:summary}, solutions in this category (\cite{lu2020blockchain}, \cite{chai2020hierarchical}, and \cite{ayaz2021blockchain}) are summarized as well.

\subsection{Federated Learning and Hierarchical Blockchain-based Systems with Centralized Aggregation (FLHB-CA)}

A centralized aggregator with a hierarchical blockchain is utilized in this system model. In \cite{lu2020blockchain}, a hierarchical structure consisting of permissioned blockchain used by Roadside Units (RSUs) and local Directed Acyclic Graphs (DAGs) used by vehicles was proposed. The DAG can be seen as the first level of the hierarchy, while the permissioned blockchain represents the upper level. Local aggregation is done between nearby vehicles, then global aggregation by RSUs. Final aggregation and model update are done by a server, which also initializes the model. The blockchain plays a role in storing the local models and global ones in addition to functioning as an authorization step since it is permissioned. A depiction of this model is given in Fig. 2(a).

\subsection{Federated Learning and Hierarchical Blockchain-based Systems with Decentralized Aggregation (FLHB-DA)}

A hierarchical blockchain is used in this system model along with decentralized aggregation as illustrated in Fig. 2(b). 
In \cite{chai2020hierarchical}, a blockchain and federated learning-based system was proposed for knowledge sharing between vehicles while maintaining privacy and security. The blockchain is hierarchical where the system has a top-chain level with one blockchain and multiple ground chains. Vehicles collect data and train a machine learning model, then send the trained parameters to an RSU which does local aggregation and stores the aggregated parameters in its ground chain. Multiple ground chains exist in different places and from each chain, a leader RSU is selected to transfer the locally aggregated parameters to the top chain. In the top chain, aggregating the parameters from different ground chains is done and the model is updated. After that, models in ground chains are updated using the global model. The evaluation was done using 5 roadside units and 6 vehicles along with MNIST and CIFAR10 datasets. About 10\% enhancement in accuracy was observed in comparison to the traditional federated learning model.

A system for message dissemination in vehicular networks was developed in \cite{ayaz2021blockchain}. A hierarchical blockchain with federated learning was used in the proposed system. The purpose of using federated learning was to prevent the need for a central server to do the training. The blockchain was used to store local and global models. Smart contracts were used to do security checks along with aggregating the parameters. Finally, the consensus protocol of the blockchain was modified to ensure choosing the next relay and disseminate the message. The trained model’s purpose is multi-hop relay selection. According to the researcher, the purpose of using a hierarchical blockchain instead of a flat one is that the hierarchical blockchain can deal with the forking problem that could happen when a vehicle is offline. That is why each vehicle stores its locally trained parameters in a microblock, where we can consider the microblock as the first level in the hierarchical blockchain. When a vehicle gets near an RSU, the RSU reads the local models, does global aggregation, and updates the global model in the main blockchain, the second level in the hierarchy.

\section{COMPARISON AND KEY FINDINGS}

The Internet of Energy domains observed in the solutions that fall into the first system model (FLB-CA) are Distributed energy trading and sharing, and Electric and connected vehicle management. Solutions based on the second system model (FLB-DA) are proposed for the domains of Smart microgrid energy networks and Electric and connected vehicle management. For the third (FLHB-CA) and fourth (FLHB-DA) system models, the solutions in these categories are proposed for the Electric and connected vehicle management domain.

The system models discussed in this paper have utilized blockchain technology for various purposes, including but not limited to:
\begin{itemize}
\item Storing data is the most basic and straightforward usage of blockchain technology. It can involve storing various types of data, including information about the peers, transactions between the peers, or the trained models.
\item Enabling peer-to-peer operations: such as energy requests, data sharing, and payments through the blockchain.
\item Replacing the centralized aggregator with smart contracts, which has been demonstrated in some of the system models.
\item Authorizing requests, which can be facilitated through permissioned or private blockchains.
\item Enhancing privacy, since blockchain operations are anonymous.
\item Validating input, particularly the trained model, through the consensus protocol of the blockchain.
\end{itemize}

For the blockchain technology aspects, we observe two types of usage as follows:
\begin{itemize}
\item Smart Contracts-based systems: All of the operations are handled by smart contracts. For example, \cite{bouachir2022federatedgrids} allows energy trading/sharing requests and prediction of future consumption of energy over smart contracts. Since smart contracts run over the blockchain, the system is blockchain-based, but we refer to that as smart-contract-based to distinguish it from the second type.

\item Blockchain-based Systems: Some researchers used not only smart contracts, but the blockchain itself as a function in the system. For example, \cite{otoum2020blockchain} used the blockchain to increase privacy since it has authentication. Other researchers modified the consensus protocol, which is a main component of a blockchain, to run some operations. For example, \cite{ayaz2021blockchain} did validation of local data and served requests through smart contracts, and the validation of trained models was done through the designed consensus protocol.
\end{itemize}

The main benefit of federated learning is providing privacy by training a global model on local peers’ data without sharing the data directly. Moreover, less bandwidth is required to share the trained parameters instead of raw data, which leads to lower latency. There is no need for a high-computing server to perform the training in federated learning, which is another significant advantage.

Regarding the level of decentralization, four levels of decentralization, namely low, medium, high, and extreme, are identified, where the more decentralized a system, the more complex it is. The decentralization level descriptions are as follows:

\textit{Low}: There is no peer-to-peer communication, and all processes are centralized, i.e. no decentralization at all, which is out of the scope of this paper.

\textit{Medium}: Training of the model is decentralized but the aggregation of the global model is centralized, which applies to system models of Federated Learning and Flat Blockchain-based Systems with a Centralized Aggregation and Federated Learning and Hierarchical Blockchain-based Systems with Centralized Aggregation.

\textit{High}: Training of the model is decentralized, and the aggregation of the global model is decentralized, but using special nodes as noticed in system models Federated Learning Flat Blockchain-based Systems with Decentralized Aggregation and Federated Learning Hierarchical Blockchain-based Systems with Decentralized Aggregation.

\textit{Extreme}: Training of the model is decentralized and the aggregation of the global model is decentralized where any peer can do it, which is also under the scope of system models Federated Learning and Flat Blockchain-based Systems with Decentralized Aggregation and Federated Learning and Hierarchical Blockchain-based Systems with Decentralized Aggregation.

As mentioned before, two different usages of blockchain were noticed. Having a system that depends only on smart contracts is useful since it can be deployed on any network, and it might be more secure against attacks to do so. However, deploying smart contracts on any network has the disadvantage of being subject to the fees of that network. Deploying on a customized network also has advantages and disadvantages. One advantage is the ability to modify the consensus protocol and use it to do certain operations as we discussed before. Another advantage is that the fees of the network would not be affected by other factors. The disadvantages are security aspects especially when the network is new or when modifying the consensus protocol without making experiments to assure its effectiveness against attacks. One possible disadvantage could be encouraging nodes to run consensus protocol and do validating since the network is new and its token would not be trusted or valuable at the beginning.

\section{OPEN ISSUES AND FUTURE WORK}

As a result of our analysis, we identify several open issues and future directions for integrating blockchains and federated learning into the Internet of Energy field. These include:

\begin{itemize}
\item Further research and development are needed to fully explore the potential of blockchain and federated learning in the fields of distributed energy trading and sharing and smart microgrid energy networks.
\item Comparing the scalability of different system models is an important area of study. This would allow researchers to identify the most suitable model for different applications and understand the performance of each model in different scenarios.
\item Developing a hierarchical blockchain and federated learning-enabled platform for energy trading and sharing is a promising area of future research. This could lead to the creation of a new and innovative platform for decentralized energy exchange.
\item One important open issue is the decision of whether to use a blockchain that is specifically designed for a particular platform or to utilize smart contracts on any blockchain, such as Ethereum \cite{wood2014ethereum} or LightChain \cite{hassanzadeh2021lightchain}. Future research needs to explore this decision-making process, as it could potentially enable researchers to determine whether it is necessary to develop their own blockchain or if they can simply use an existing one that supports smart contracts for their platform.
\item Cyber-attacks affect such systems more since they will affect selling and buying energy or shared knowledge. Protection methods and the development of secure and upgradeable smart contracts are important areas of research in this field.
\item Explore the potential of DHT-based blockchains\cite{zhang2013distributed, abe2018mitigating, hassanzadeh2021lightchain, wu2021mapchain, chen2022scalable} integrated with federated learning for Internet of Energy systems, as no current solution has been proposed in this area. Such a system model could provide significant benefits for the Internet of Energy.
\end{itemize}

\section{CONCLUSION}

In this paper, we have focused on the Internet of Energy and its major domains. Key technologies relevant to the Internet of Energy, including flat blockchains, hierarchical blockchains, smart contracts, and distributed federated learning, have been defined. We have presented four system models that have been extracted from state-of-the-art solutions which have utilized blockchain and federated learning in the context of the Internet of Energy. These system models demonstrate potential ways for integrating blockchain and federated learning technologies in the various domains of the Internet of Energy.

The integration of federated learning with decentralized aggregation in both flat and hierarchical blockchain-based systems has been shown to achieve higher levels of decentralization compared to the counterparts using centralized aggregation. The hierarchical model, with its scalable and low-latency structure, appears to be the most favorable option among the decentralized models. Conversely, the flat model demonstrates lower decentralization and efficiency.

\section*{Acknowledgment}
\noindent This work is supported by TUBITAK (The Scientific and Technical Research Council of Türkiye) 2247-A National Leader Researchers Award 121C338. A very preliminary version of this work was accepted for poster presentation at IEEE ICBC conference \cite{azekiye}.

\bibliographystyle{IEEEtran}
\bibliography{IEEEabrv,paper}

\end{document}